\documentstyle[12pt]{article}
\begin{document}
\title{Comment on "Hara's theorem in the constituent quark model"}
\author{
{P. \.{Z}enczykowski}$^*$\\
\\
{\em Dept. of Theor. Physics} \\
{\em Institute of Nuclear Physics}\\
{\em Radzikowskiego 152,
31-342 Krak\'ow, Poland}\\
}
\maketitle
\begin{abstract}
It is pointed out that current conservation {\em alone} does not suffice 
to prove Hara's
theorem as it was claimed recently.
By explicit calculation we show that the additional implicit assumption
made in such "proofs" is that of a sufficiently localized current.

\end{abstract}
$^*$ E-mail:zenczyko@solaris.ifj.edu.pl\\
\vskip 0.8in
\begin{center}REPORT \# 1765/PH, INP-Krak\'ow \end{center}
\newpage

Weak radiative hyperon decays proved to be a challenge to our theoretical
understanding.  Despite many years of theoretical studies a
satisfactory description of these processes is still lacking.
In a recent review (see ref.\cite{LZ}) presenting the current
theoretical and experimental situation in that
field, attention was focussed on the question of the validity of
Hara's theorem \cite{Hara}.  
This question was originally 
posed by the paper of Kamal and Riazuddin \cite{KR} who observed that in the
quark model Hara's theorem is violated. 
There have been several attempts to understand the origin and meaning of this 
quark model result 
\cite{LiLiu,Zen,Dmi96,Azimov}.
Here we want to comment on ref. \cite{Dmi96} wherein it is claimed that
the argument made by Serot in ref. \cite{Serot} and discussed later in
ref.\cite{Dmi92} is sufficient to prove Hara's theorem. 

In ref.\cite{Dmi96} it is stated that the argument of Serot (upon which the
claim of ref.\cite{Dmi96} is based) relies {\em only}
on the multipole decomposition
of the electromagnetic current matrix element and on 
the conservation of electromagnetic current.
This statement should be treated with suspicion as in the standard
proof of Hara's theorem it is the absence of massless hadrons that 
- besides gauge invariance - is necessary
for the proof to go through (see, eg. ref.\cite{LZ}). 
Thus, one may suspect that the argument of Serot uses 
a somewhat similar additional {\em hidden}
assumption.
Rather than trying to identify such an implicit assumption, a large part
of ref.\cite{Dmi96} (see also ref.\cite{Dmi92}) is then concerned with the
demonstration of how to satisfy electromagnetic current conservation 
in actual calculations with composite states. 
Below we will demonstrate through an explicit calculation 
what implicit assumption is being made
in Serot-like arguments. 

There are two conserved electromagnetic 
currents entering into the discussion of Hara's theorem:
\begin{equation}
\label{eq:gamma5}
J_5^{\mu }= F_1(q^2) \overline{\psi _1}
(\gamma ^{\mu } - \frac{q^{\nu }\gamma_{\nu}}{q^2} q^{\mu })\gamma _5 \psi _2 
\end{equation}
and
\begin{equation}
\label{eq:sigma5}
J_5^{'\mu }= F_3(q^2)\overline{\psi _1} 
i \sigma ^{\mu \nu } q_{\nu }\gamma _5 \psi _2
\end{equation}

In the limit of exact SU(3) ($m_1 = m_2$) 
the coupling of photon to current $J'_5$  
vanishes due to its symmetry properties (see Sec. 3.1. in ref.\cite{LZ}).
The only allowed current is then that of Eq.(\ref{eq:gamma5}). According to 
Zeldovich and Perelomov \cite{Zel61} expansion of $F_1(q^2)$ around $q^2 = 0$ 
has to start with a term proportional to $q^2$. For a real photon this entails 
a vanishing current matrix element.
However, as discussed in ref.\cite{Dmi92} for a nonvanishing $F_1(0)$
one obtains current 
matrix element which is finite at $q^2=0$ and a vanishing parity-violating
charge density. Thus, the form of Eq.(\ref{eq:gamma5})
seems fully admissible also for $F_1(0) \ne 0$.
In ref.\cite{Dmi92} it is then claimed that Serot managed to prove the 
vanishing of the relevant matrix element at $q^2=0$ using 
conservation of the electromagnetic current {\em only}. 
As remarked above the proof
of Serot most likely uses a hidden assumption.
Let us therefore look at this proof in some detail.

In the nonrelativistic approximation the current $J_5^{\mu }$ 
takes the form (see also
ref.\cite{Dmi92}):
\begin{equation}
\label{eq:J5nonrel}
{\bf J}_5 ({\em q})=
\frac{\bf{q}\times(\mbox{\boldmath{$\sigma $}}\times \bf{q})}{\bf{q}^2}= 
\mbox{\boldmath {$\sigma $}}
- (\mbox{\boldmath {$\sigma $}} \cdot {\bf \hat{q}})\,{\bf \hat{q}}  
\end{equation} 
where ${\bf \hat{q}} = {\bf q}/|{\bf q}|$ and we have put $F_1(0)=1$ .
For ${\bf J}_5(q)$ of Eq.(\ref{eq:J5nonrel}) the transverse
electric dipole is clearly nonzero.

On the other hand, the argument of
Serot, which starts with a general formula for the transverse electric dipole,
seems to show that for 
${\bf q}^2 \rightarrow 0$ this multipole vanishes as ${\bf q}^2$ anyway.
Since the argument of Serot is made in position space,
in order to analyze it we have to find the 
shape of current ${\bf J}_5$ from Eq.(\ref{eq:J5nonrel}) in position space.
Let us therefore consider
\begin{equation}
\label{eq:Fourier}
{\bf J}^{\varepsilon}_5 ({\bf r})= \frac{1}{(2\pi )^3}\int d^3q \;
 {\bf J}({\bf q})\,
e^{-i{\bf q}{\bf r}- \varepsilon {\bf q}^2}
\end{equation}
In Eq.(\ref{eq:Fourier}) we have introduced a small parameter ($\varepsilon $) to 
regularize the emerging integrals.

The integral in Eq.(\ref{eq:Fourier}) is composed of two pieces. The first
(Fourier transform of 
${\bf J}^{(1)\varepsilon }({\bf q}) \equiv \mbox{\boldmath{$\sigma $}} 
\exp (-\varepsilon {\bf q}^2)$) gives:
\begin{equation}
\label{eq:J(1)}
{\bf J}^{(1) \varepsilon }({\bf r}) = \mbox{\boldmath{$\sigma $}}
\cdot \delta _{\varepsilon }^3({\bf r})
\end{equation}
with
$\delta _{\varepsilon }^3({\bf r})=
\prod _{i=1}^3 \delta _{\varepsilon }(r_i)$, where
$\delta _{\varepsilon }(r_i)$ is a one-dimensional 
regularized delta function:
\begin{equation}
\label{eq:delta}
\delta _{\varepsilon }(r_i)=\frac{1}{\sqrt{4\pi \varepsilon}} 
\exp (-r_i^2/(4 \varepsilon ))
\end{equation}

Calculation of the second piece (Fourier transform of the term
${\bf J}^{(2)\varepsilon }({\bf q}) \equiv
-(\mbox{\boldmath{$\sigma $}}\cdot{\hat{\bf q}})\,{\hat{\bf q}}
\exp (-\varepsilon {\bf q}^2)$) is more
complicated and is sketched in Appendix.
Together one obtains:
\begin{eqnarray}
\label{eq:fullJ(r)}
{\bf J}_5^{\varepsilon }({\bf r}) & = &
{\bf J}_5^{(1)\varepsilon }({\bf r}) + {\bf J}_5^{(2)\varepsilon }({\bf r}) 
\nonumber \\
&=&[\mbox{\boldmath{$\sigma $}}-
(\mbox{\boldmath{$\sigma $}}\cdot \hat{\bf r})\,\hat{\bf r}]
\,\delta _{\varepsilon }^3({\bf r})  
+\frac{1}{2\pi r^2}\,
[\mbox{\boldmath{$\sigma $}}-
3(\mbox{\boldmath{$\sigma $}}\cdot \hat{\bf r})\,\hat{\bf r}]
\,\delta _{\varepsilon}(r) 
\nonumber \\&&-
\frac{1}{4\pi r^3}\,
[\mbox{\boldmath{$\sigma $}}-
3(\mbox{\boldmath{$\sigma $}}\cdot \hat{\bf r})\,\hat{\bf r}]
\: {\rm erf} \left(\frac{r}{2\sqrt{\varepsilon }}\right)
\end{eqnarray}
where
${\rm erf} (x) = \frac{2}{\sqrt{\pi }}\int_{0}^{x}e^{-t^2}\,dt$ 
is the error function,
$\hat{\bf r}={\bf r}/r$, and $r=|{\bf r}|$.

The transverse electric dipole is defined as \cite{Dmi96,Dmi92,FW}: 
\begin{equation}
\label{eq:T1el}
T^{el}_{1M}= \frac{1}{i q \sqrt{2}} \int d^3r \left[ 
-{\bf q}^2 ({\bf J}_5\cdot {\bf r})+
(\nabla \cdot {\bf J}_5)[1+({\bf r}\cdot \nabla)]
\right]
\,j_1(qr) \,Y_{1M}({\bf \hat{r}})
\end{equation}
where $q = |{\bf q}|$.

One may check by direct calculation that current ${\bf J}_5^{\varepsilon }(
{\bf r})$ of Eq.(\ref{eq:fullJ(r)}) is conserved
(as it should be because for its Fourier transform ${\bf J}_5
^{\varepsilon }({\bf q})$
we obviously have from Eq.(\ref{eq:J5nonrel}): 
${\bf q}\cdot {\bf J}_5^{\varepsilon }({\bf q}) =0$ ):
\begin{eqnarray}
\label{eq:Jcons}
\nabla \cdot {\bf J}_5^{\varepsilon } &=& 
\left(-\frac{2 \mbox{\boldmath {$\sigma$}}\cdot {\bf \hat{r}}}{r} 
\,\delta ^3_{\varepsilon }({\bf r})\right) +
\left(\frac{2 \mbox{\boldmath {$\sigma$}}\cdot {\bf \hat{r}}}{r} 
\,\delta ^3_{\varepsilon }({\bf r})-
\frac{\mbox{\boldmath {$\sigma$}}\cdot {\bf \hat{r}}}{\pi r^3}
\,\delta _{\varepsilon }(r)\right)
+\left(\frac{\mbox{\boldmath {$\sigma$}}\cdot {\bf \hat{r}}}{\pi r^3}
\,\delta _{\varepsilon }(r)\right)
\nonumber \\
& = & 0
\end{eqnarray}
In Eq.(\ref{eq:Jcons}) the three terms in parantheses come from the three 
terms on the right-hand side of Eq.(\ref{eq:fullJ(r)}).
Clearly, all three terms in Eq.(\ref{eq:fullJ(r)}) are required for
the cancellation of Eq.(\ref{eq:Jcons}) to work.
Since $\nabla \cdot {\bf J}_5^{\varepsilon }=0$, 
in the calculation of the electric dipole in Eq.(\ref{eq:T1el}) 
only the first term of the integrand may
give a nonzero result.
However, according to the argument of ref.\cite{Dmi96,Dmi92} 
for small $q$ this term is proportional to $q^2$ after replacing the 
spherical Bessel
function $j_1(qr)$ by its approximation for small arguments: $\frac{1}{3}qr$.
Consequently, the argument seems to show that $T^{el}_{1M}$ vanishes in the
long wavelength limit
$q^2 \rightarrow 0$.

Unfortunately, the above argument is not correct when one uses current
${\bf J}_5^{\varepsilon }$ of Eq.(\ref{eq:fullJ(r)}). 
Let us calculate:
\begin{equation}
\label{eq:J.r}
{\bf J}_5^{\varepsilon }({\bf r}) \cdot {\bf r} = 
-\frac{\mbox{\boldmath {$\sigma$}}\cdot {\bf \hat{r}}}{\pi r}
\;\delta _{\varepsilon }(r)
+ \frac{1}{2 \pi r^2} \;\mbox{\boldmath {$\sigma$}} \cdot {\bf \hat{r}} 
\;\,{\rm erf} \left( \frac{r}{2 \sqrt{\varepsilon }}\right)
\end{equation} 
The first term on the right-hand
side of Eq.(\ref{eq:J.r}), being proportional to delta function, 
is sufficiently localized in space to permit the replacement of the
spherical Bessel function $j_1$ by its approximation for small arguments, since
only small values of $qr$ are allowed ($q \rightarrow 0$ and $r$ is small).  
With the second term the situation 
is, however, different.

Introducing new variable $z \equiv qr$ the contribution of the second term
in Eq.(\ref{eq:J.r}) to the transverse electric dipole is
(for $\varepsilon \rightarrow 0$)
\begin{equation}
\label{eq:T1elz}
T^{el}_{1M}= \frac{i}{2 \pi \sqrt{2}}
\int_{0}^{\infty} dz \,j_1(z) 
\int d\Omega _{\hat{z}}\;\mbox{\boldmath {$\sigma$}}\cdot {\bf \hat{z}}
\;Y_{1M}({\bf \hat{z}})
\end{equation}
The right hand side apparently does not depend on $q$. 
Using $\frac{d}{dz}\; j_0(z) = -j_1(z)$ and 
$Y_{1M}({\hat{\bf z}})=
\sqrt{\frac{3}{4\pi}}\mbox{\boldmath{$\epsilon $}}_M \cdot{\hat{\bf z}}$
we calculate
\begin{equation}
\label{eq:finalT1M}
T^{el}_{1M}=\frac{i}{\sqrt{6\pi}}
\mbox{\boldmath{$\epsilon $}}_M \cdot \mbox{\boldmath {$\sigma$}}
\end{equation}
a definitely nonvanishing result in agreement with Eq.(\ref{eq:J5nonrel}).
The origin of this nonzero result is  clear
from the above calculation: in Eq.(\ref{eq:fullJ(r)}) 
the third term of axial current 
${\bf J}_5^{\varepsilon}({\bf r})$ is {\em not}
localized in space sufficiently well.
Thus, the theorem of Serot is based not only on current conservation but
also on the assumption that the position-space current vanishes at infinity
faster than $1/r^3$.
To forbid such a behaviour corresponds in standard proofs of Hara's theorem
to assuming the absence of massless (infinite range) hadrons.

It is a different question whether the above-identified 
implicit assumption used by Serot
should really be made. Conventional wisdom certainly requires the 
electromagnetic axial current of a baryon to be well localized in position
space, which assumption - together with that of current conservation - 
leads to a vanishing parity-violating matrix element of the electromagnetic
current at the real photon point.

Results of strict quark model calculations of Kamal and Riazuddin (KR) 
\cite{KR}, which indicated the nonvanishing of this matrix element 
(for $SU(3)$-related strangeness-changing current 
$\Sigma ^+ \rightarrow p$), were therefore treated with disbelief. 
However, since Serot theorem is not based on current conservation {\em only}, 
one cannot 
conclude from the violation of Hara's theorem obtained in ref.\cite{KR}
that gauge invariance must be broken in these calculations. In fact,
by repeating KR calculations one can convince oneself that gauge invariance 
{\em is preserved} in ref.\cite{KR}. 
Thus, it seems that it is rather the other assumption used by Serot: that of
a sufficiently well localized current, which is violated in the KR paper.

This tentative identification seems understandable if one
thinks of quark model prescription in position space. Indeed, in the strict
quark model, the initial and final states are described by sums of 
tensor products of plane-wave quark states spreading all over position space.
In the calculation of KR 
the intermediate quark (between the action of weak Hamiltonian and the 
emission of a photon) may also propagate to spatial infinity, 
reflecting total quark freedom.
It should not come then as 
a surprise that the total electromagnetic current
of the three-quark state
contains a piece which is not sufficiently well localized.

To summarize let us repeat: the assumption of 
current conservation {\em alone}
does not suffice to prove Hara's theorem. 
The current of Eq.(\ref{eq:fullJ(r)}) is definitely conserved and
yet the transverse electric dipole moment is nonzero.
Thus, the considerations of refs.\cite{Dmi96,Dmi92} which concern
the detailed manner in which current conservation is realized for
composite systems cannot by themselves provide us with a proof of
Hara's theorem.

ACKNOWLEDGEMENTS.

This work was partially supported by the
KBN grant No 2P0B23108.
Discussions with M. Sadzikowski are gratefully acknowledged.

\newpage

APPENDIX

Calculation of ${\bf J}^{(2)\varepsilon}({\bf r})$ requires determination
of the integral
\begin{equation}
\label{eq:A1}
I_{ml}({\bf r}) = - \frac{1}{(2\pi)^3}\int d^3q \; \frac{q_mq_l}{q^2}
\; e^{-i{\bf q}\cdot{\bf r}-\varepsilon {\bf q}^2}
\end{equation}
which may be evaluated as
\begin{eqnarray}
\label{eq:A2}
I_{ml}({\bf r}) &= &\frac{1}{(2\pi)^3} \frac{\partial ^2}
{\partial r^m\,\partial r^l}
\int _{\varepsilon }^{\infty} d\xi \int d^3q \;
e^{-\xi {\bf q}^2-i{\bf q}\cdot{\bf r}} \nonumber \\
&=&\frac{1}{4{\pi}^{3/2}}\frac{\partial ^2}{\partial r^m\,\partial r^l}\;
\frac{\sqrt{\pi}}{r}\;{\rm erf} \left(\frac{r}{2\sqrt{\varepsilon}}\right)
\end{eqnarray}
Performing indicated differentiations we obtain
\begin{eqnarray}
\label{eq:A3}
I_{ml}({\bf r}) &= &-\frac{1}{4\pi r^3}(\delta _{ml}-3\hat{r}_m\hat{r}_l)\;
{\rm erf}\left(\frac{r}{2\sqrt{\varepsilon}}\right) \nonumber \\
&&
-\frac{1}{\pi r^2}\; \hat{r}_m\hat{r}_l \; \frac{1}{\sqrt{4\pi \varepsilon}}
\exp \left(-\frac{r^2}{4 \epsilon }\right)\nonumber \\
&&
+
\frac{1}{2 \pi r^2}(\delta _{ml}-\hat{r}_m\hat{r}_l)
\frac{1}{\sqrt{4\pi \varepsilon}}
\exp \left(-\frac{r^2}{4 \epsilon }\right)
\nonumber \\
&&
-\hat{r}_m\hat{r}_l \;\frac{1}{(4\pi \varepsilon )^{3/2}}
\exp \left(-\frac{r^2}{4 \epsilon }\right) 
\end{eqnarray}
This leads to Eq.(\ref{eq:fullJ(r)}).


\end{document}